\documentclass[aps,prd,reprint,twocolumn,preprintnumbers,nofootinbib]{revtex4}
\usepackage{amsmath}
\numberwithin{equation}{section}
\usepackage{amssymb}
\usepackage{amsfonts}
\usepackage{hyperref}
\usepackage{epsfig}
\usepackage{graphicx,bm}
\usepackage{xcolor}
\usepackage{slashed} 
\usepackage{mathrsfs} 
\def\beq{\begin{equation}}
\def\eeq{\end{equation}}
\def\bea{\begin{eqnarray}}
\def\eea{\end{eqnarray}}
\def\ba{\begin{array}}
\def\ea{\end{array}}
\def\l{\lambda}

\def\t1{\tilde{t}_1}
\def\t2{\tilde{t}_2}
\def\b1{\tilde{b}_1}

\newcommand{\bd}{\begin{displaymath}}
\newcommand{\ed}{\end{displaymath}}
\newcommand{\be}{\begin{equation}}
\newcommand{\ee}{\end{equation}}

\def\b{\beta}

\def\l{\lambda}

\def\q2 {q^2}

\def\t {\times }

\def\bt{\begin{table}}
\def\et{\end{table}}

\catcode`@=11 
\def \gsim{\mathrel{\mathpalette\@versim>}}
\def \lsim{\mathrel{\mathpalette\@versim<}}
\def \@versim#1#2{\lower0.4ex\vbox{\baselineskip\z@skip\lineskip\z@skip
     \lineskiplimit\z@\ialign{$\m@th#1\hfil##\hfil$%
     \crcr#2\crcr\sim\crcr}}}
\catcode`@=12 

\begin{document}

\title{%
Diphoton signal via Chern-Simons interaction in a warped geometry scenario}%

	
\author{Nabarun Chakrabarty\footnote{nabarunc@hri.res.in} and Biswarup Mukhopadhyaya\footnote{biswarup@hri.res.in}} \affiliation{Regional Centre for Accelerator-based Particle Physics, \\
     Harish-Chandra Research Institute,Chhatnag Road, Jhusi, Allahabad 211019, India}

\author{Soumitra SenGupta\footnote{tpssg@iacs.res.in}} \affiliation{Department of Theoretical Physics,
  Indian Association for the Cultivation of Science, 2A $\&$ 2B Raja
  S.C. Mullick Road, Kolkata 700032, India}

\begin{abstract}
\begin{center}
{Abstract}
\end{center}

\vspace{0.1cm}
The Kalb-Ramond field, identifiable with bulk torsion in a 5-dimensional
Randall Sundrum (RS) scenario, has Chern-Simons interactions with gauge bosons, from the requirement of gauge anomaly cancellation. Its lowest 
Kaluza Klein (KK) mode on the visible 3-brane can be identified with a spin-0 CP-odd field, 
namely, the axion.
By virtue of the warped geometry and Chern-Simons couplings, this axion has 
unsuppressed interactions with gauge bosons in contrast to ultra-suppressed interactions with fermions.
The ensuing dynamics can lead to a peak in the diphoton spectrum, which could  
be observed at the LHC, subject to the prominence of the signal. 
Moreover, the results can be numerically justified when the warp factor
is precisely in the range required for stabilisation of the electroweak scale. 

\end{abstract}
 
\maketitle

\section{Introduction}

Suppose there is fundamental spin-0 particle in nature, whose couplings to all 
Standard Model (SM) particles are extremely suppressed,
the only exception being pairs of gauge bosons. In such a case, such interactions should constitute the sum and substance of its phenomenology observable at the Large Hadron Collider (LHC), provided that it is within the kinematic reach of the latter. 

Such a situation
is not altogether far-fetched. A case in point is a CP-odd spin-0 axion field, in terms of which 
a Kalb-Ramond (KR) antisymmetric tensor field strength can be defined. 
As we shall emphasize further in the remaining part of this paper, the KR field exhibits 
some very interesting properties if it propagates in bulk in a (1+4) dimensional warped geometry scenario as proposed first by Randall and Sundrum. However, the theory is not in general anomaly-free if it arises from a still higher-dimensional scenario such as 10-dimensional supergravity. 
This problem is avoided (see section \ref{sec:scenario})
if the KR field is endowed with Chern-Simons (CS) terms couplings
with gauge bosons, also propagating in the bulk~\cite{GREEN1984117}. On compactification of the warped extra dimension, the CS term has unsuppressed interaction strengths 
of its zero mode with gauge boson pairs on the (1+3) dimensional visible brane. This unsuppressed character is in turn translated into the interaction of the axion 
field, in terms of which the zero-mode KR field strength is expressed. On the other hand, the fermionic couplings of the CS-axion turn out to be suppressed as an artefact of the given scenario.
With a non-perturbatively acquired mass of this
axion, it can have interesting LHC phenomenology where the CS-driven gluon fusion process can produce it, followed by its decays into gauge boson pairs, of which the most spectacular signal consists in diphoton invariant mass peaks. The expected rates of such peaks are estimated in this paper. We mention in this connection that a recent surge of interest on such diphoton peaks came with the apparent 
occurrence of a diphoton peak at about
750 GeV in the initial 13 TeV run of the LHC. It was first reported in 
the preliminary announcement of the 13 TeV run \cite{atlas:15dec,CMS:2015dxe} and  
corroborated in the recent reports in the recent Moriond meeting~\cite{atlas:morion}. 
It led to an
avalanche of explanations offered in context of various new physics scenarios(see for a representative list
~\cite{Staub:2016dxq,PhysRevD.93.055027,Nakai:2015ptz,Arcadi:2016dbl,Dev2016,
Bardhan:2016rsb,Perelstein:2016cxy,
Bernon:2016dow,Chiang:2016eav,
Ellwanger:2016qax,PhysRevD.93.055025,Chakrabortty:2015hff,
Falkowski2016,PhysRevD.93.055034,
PhysRevD.93.055011,Angelescu2016126,Gabrielli201636,Harigaya2016151,
Mambrini2016426,Buttazzo2016,Heckman2016231,PhysRevD.93.055006,Salvio2016469,Anchordoqui2016312,
Nomura2016306,Aydemir:2016qqj,King2016,Kim2016403,PhysRevD.93.035002}). Even though the signal ultimately failed to persist, it could nonetheless glorify the importance of the diphoton final state in detecting a spin-0 TeV-scale resonance, that could indeed be a reality for higher
axion masses.

As has
been stated already, we present our study in the context of
a five-dimensional Randall-Sundrum scenario with bulk
space-time torsion identifiable with a Kalb-Ramond tensor field. 
The massless four-dimensional projection of this
field is expressible in terms of an axion which may acquire
a mass term non-perturbatively. While this axion field
has extremely suppressed coupling with fermions on the
four-dimensional visible brane, it can have enhanced 
interaction with gauge boson pairs via Chern-Simons(CS)
terms. Using such CS terms as the driving dynamics, we examine
the diphoton production rates in this study.  
To make matters more practical from the experimenatal perspective, we
choose only those values of the warp factor in the five-dimensional geometry, 
which not only can explain the hierarchy between the Planck
and electroweak scales, but also generate KK gauge boson masses
above the lower bound set by LHC. We present our results for different axion masses.

This paper is planned as follows. In Section~\ref{sec:scenario}, we introduce the
theoretical framework. Prospects of embedding the recently observed resonance
in that framework are studied in Section~\ref{750}. A more general discussion
on other possible diphoton resonances can be found in Section~\ref{750}.
We summarise in Section ~\ref{summary}.

\section{The scenario}\label{sec:scenario}
As stated at the beginning, we consider a five-dimensional Randall-Sundrum (RS) scenario, 
where the extra dimension is a $\mathcal{Z}_2$
orbifold of radius $r_c$. There are two branes at the orbifold fixed points i.e., 
$\phi = 0$ and $\phi = \pi$, where $\phi$ is the angular variable for the compact co-ordinate. 
The five-dimensional metric is
\bea
ds^2 &=& e^{-2 k r_c |\phi|} \eta_{\mu \nu} dx^{\mu}dx^{\nu} + r_c^2 d\phi^2 
\eea
with $\eta_{\mu \nu} = \{-,+,+,+\}$ is the Minkowski metric and $k$ (related to the 
bulk cosmological constant)
is of the order of the  four-dimensional Planck mass $M_{P}$.

The five-dimensional Planck mass $M$ is related to $M_P$ as,
\bea
M^2_P &=& \frac{M^3}{k}(1 - e^{-2 k r_c \pi})
\eea

Resolution of the naturalness problem requires
$k r_c \simeq 11.5$. Attempts towards finding a warped-geometric explanation to the 750 GeV 
resonance can be seen in
\cite{Carmona:2016jhr,Hewett:2016omf,Csaki:2016kqr,Falkowski:2016glr,Bauer:2016lbe,Sanz:2016auj,
Giddings:2016sfr,Martini:2016ahj,Cai:2015hzc,Ahmed:2015uqt,Arun:2015ubr,Cox:2015ckc,
Dillon:2016fgw,Bardhan:2015hcr}

The presence of CS terms can be motivated if the aforesaid scenario is the 
descendant of, say, a ten dimensional supergravity theory. There the KR field strength 
tensor is augmented by a set of CS terms
in order to cancel gauge anomalies\footnote{The massless sector of D =10 supergravity multiplet contains a second rank antisymmetric tensor field $B_{\mu\nu}$ with a
corresponding third rank field strength $H_{\mu\nu\alpha}$. This field
strength can can also be interpreted as  the background space-time
torsion. The field $B_{\mu\nu} $ plays a crucial role in canceling the
gauge anomaly originating from the one loop hexagon diagrams with six
external legs of gauge fields with chiral fermions in the loop. If the
third rank field strength is now  modified with appropriate Chern-Simons 
term as $H_{\mu\nu\alpha} = \partial_{[\mu} B_{\nu\alpha ]} +
\frac{A_{[\mu} F_{\nu\alpha]}}{M_P}$, then the corresponding tree diagramm
that is generated with Kalb-Ramond field as propagator between the gauge
fields  at the two vertices exactly  cancels the hexagon gauge anomaly~\cite{GREEN1984117}.}.
This anomaly cancellation feature is protected in the 4-dimensional
effective theory with a modification of the  gauge-KR coupling by an
appropriate volume modulus factor.

When the five-dimensional space-time has torsion along with curvature, the torsion field
can be identified with the KR field
\cite{PhysRevD.9.2273,0264-9381-19-4-304,0264-9381-16-12-102,
PhysRevD.70.066009,PhysRevD.72.066012,PhysRevD.77.015010,
Mukhopadhyaya19998,doi:10.1142/S0217732302006151,0264-9381-21-14-004,PhysRevLett.89.121101}.
It has been shown earlier that such a field is suppressed in (1+3) dimensions due
to the RS geometry, thus explaining why our observed universe is controlled primarily
by curvature rather than torsion.

The five-dimensional Einstein-Maxwell-Kalb-Ramond (EKMR) action in the Einstein frame reads
\bea
S_{eff} &=& \int d^5 x \sqrt{-G}\Big[R - \frac{1}{4}F_{MN}F^{MN}\\ \nonumber
&&
 - \frac{1}{12}\overline{H}_{MNL} \overline{H}^{MNL}\Big] \\
\overline{H}_{MNL} &=& H_{MNL} + \frac{2}{M^{3/2}} B_{[M} F_{NL]} + \frac{2}{M^{3/2}} W^i_{[M} W^i_{NL]}\nonumber \\ 
&&
+ \frac{2}{M^{3/2}} G^b_{[M} G^b_{NL]} 
\eea

where a sum over i = 1,2,3 and b = 1,2..,8 is implied. In addition we have
\bea
H_{MNL} &=& \partial_{[M} B_{NL]}
\eea
Here $B_{NL}$ refers to the Kalb-Ramond (KR) two form in five-dimensions. Besides, $B_M(x,\phi)$, $W^i_M(x,\phi)$ and $G^b_M(x,\phi)$ respectively refer to the $U(1)_Y$, $SU(2)_L$ and $SU(3)_c$ gauge fields in the
bulk with $F_{NL}$, $W^i_{NL}$ and $G^b_{NL}$ as the corresponding field strengths \footnote{The CS coupling in 5 dimensions will always carry a $\frac{1}{M^{3/2}}$ on dimensional grounds. However, the numerical factors may not be strictly same for all
gauge fields. For simplicity, we assume universal couplings in this work.}. The gauge $SU(2)_L$ and $SU(3)_c$ gauge indices read $i$ \text{and} $b$ respectively. Further, KR gauge invariance allows us to do 
gauge fixing using $B_{\mu y} = 0$ .

With the standard model (SM) gauge fields in the bulk, the Kaluza-Klein towers
for them as well as the KR field on the visible brane are given by
\bea
B_{\mu \nu}(x,\phi) &=&\sum_{n=1}^{\infty} B^{n}_{\mu \nu}(x) \frac{\chi^{n}(\phi)}{\sqrt{r_c}} \\
C_{\mu}(x,\phi) &=& \sum_{n=1}^{\infty} C^{n}_{\mu}(x) \frac{\psi^{n}(\phi)}{\sqrt{r_c}} 
\eea
where $C$ stands for the towers corresponding to the SM gauge fields, viz, $B$, $W$ and $G$.
The zero-mode for the KR field obeys the following equation.
\bea
\frac{1}{r^2_c} \frac{d^2 \chi^0 }{d \phi^2} = 0
\eea
The solution reads
\bea
\chi^0(\phi) = c_1 + c_2 |\phi|
\eea
Continuity of the first derivative of $\chi^0(\phi)$ at the orbifold fixed
points $\phi = 0, \pm \pi$ gives $c_2 = 0$. $c_1$ is fixed using the orthonormality
condition
\bea
\int e^{2 k r_c \phi} \chi^m(\phi) \chi^n(\phi) d\phi = \delta_{m n}
\eea
This leads to the following zero-mode profiles. 
\bea
\chi^0(\phi) &=& \sqrt{2 k r_c} ~e^{- k r_c \pi} \\
\psi^0(\phi) &=& \frac{1}{\sqrt{2 \pi}} 
\eea

 $H^{0}_{\mu\nu\l}$, the field strength of $B^{0}_{\mu \nu}$ $H^{0}_{\mu\nu\l}$ can be expressed as
\bea
H^{0}_{\mu\nu\l} &=& \epsilon_{\mu\nu\l\rho}~\partial^{\rho}a \nonumber
\eea
Here $a$ denotes a CP-odd scalar, called the KR axion. 
Such an axion acquires a mass term through non-perturbative effects confined to the TeV brane
such as instanton corrections~\cite{PhysRevD.61.105007}
This mass is {\it prima facie} 
a free parameter, and which can be around a TeV scale\footnote{It should be noted, however, that the possibility of this mass lying within the reach of the LHC is \emph{not predicted} by the scenario considered here. The analysis presented by us is more in a 'looking under the lamppost' spirit, since
diphoton signals are in any case of great curiosity and phenomenological significance.}.  
The kinetic terms of $a$ and its coupling to the SM gauge fields via the Chern-Simons 
terms take the form\cite{PhysRevD.72.066012}
\bea
S_{Kin} &=& -\frac{1}{2} \eta^{\mu \nu} \partial_{\mu}a ~\partial_{\nu}a \\
S_{CS} &=& f~\Big[a B_{\mu\nu} \tilde{B}^{\mu\nu} + a W^i_{\mu\nu} \tilde{W}^{i\mu\nu} \nonumber  \\
&&
+ a G^b_{\mu\nu} \tilde{G}^{b\mu\nu}\Big] 
\eea

where, $f = -\frac{e^{k r_c \pi}}{ \sqrt{2} \pi k r_c M_{P}}$ quantifies the coupling of the axion to the SM gauge bosons. Moreover, $\tilde{B}^{\mu\nu} = \frac{1}{2} \epsilon^{\mu\nu\l\rho}~B_{\l\rho}$ ~\text{etc. denote the 
dual of the original field strength.}

It should be noted that the CS terms enable the axion to have enhanced coupling
to gauge field pairs, by virtue of the specific nature of the warped geometry.
In contrast, it has been  found~\cite{PhysRevLett.89.121101,PhysRevD.90.107901} that $a$ has interaction to
fermion pairs of the form$\sim \frac{e^{-k r_c \pi}}{M_{P}}$. As a result,
both its production rate via gluon fusion and its diphoton partial decay width
are enhanced to an extent to be decided by the acquired mass of the axion
on the TeV brane. 

The expressions for the leading order decay widths of $a$ to various $VV$ (pair of gauge-bosons) states 
are 
\bea
\Gamma_{a \rightarrow \gamma \gamma} &=& \frac{1}{4\pi} f^2 m_a^3 \\
\Gamma_{a \rightarrow g g} &=& \frac{2}{\pi} f^2 m_a^3 \\
\Gamma_{a \rightarrow W W} &=& \frac{f^2 m^3_a}{2\pi}\Big(1 - \frac{4 m^2_W}{m^2_a}\Big)^{3/2} \\
\Gamma_{a \rightarrow Z Z} &=& \frac{f^2 m^3_a}{4\pi}\Big(1 - \frac{4 m^2_Z}{m^2_a}\Big)^{3/2}
\eea

\section{Analysis strategy and numerical prediction.}\label{750}
  
We have the following expression for $a$ production cross section via gluon fusion
\bea
\sigma_{p p \rightarrow a}(fb) = c_{g g} \frac{\Gamma_{a \rightarrow g g}(GeV)}{m_a s}\times 0.3894\times 10^{12}
\eea
Here $c_{gg}$ comes from convoluting over the parton densities.
\be
c_{g g} =  \frac{\pi^2}{8} \int \frac{dx}{x} g(x) g(\frac{m^2_a}{x s})
\ee

For practical purposes, we take $c_{gg} \simeq$ 2137. The cross section
to the diphoton final state is then straightforwardedly obtained by
multiplying with the corresponding branching ratio.
\be
\sigma_{p p \rightarrow a \rightarrow \gamma \gamma} = \sigma_{p p \rightarrow a} \times \frac{\Gamma_{a \rightarrow \gamma \gamma}}{\Gamma_a}
\ee
The couplings of the
axion to the gauge boson pairs are taken to be universal in this study. This makes the branching ratio to a given $VV$ state independent of $k r_c$, for a fixed axion mass. 

Both the production cross section of $a$ as well as its partial width to $g g$ state 
are prone to QCD corrections. To encapsulate its effect, one can in principle scale both the
production cross section as well the partial width by some $K_{QCD}$. Considering that 
the dominant contribution to $\Gamma_a$ comes from the $g g$ state and that $K_{QCD}$ is 
expected to be greater than unity, its effect in the diphoton cross section 
largely cancels out.

We mention in this context that we have also implemented the effective Lagrangian
into the \texttt{FeynRules} package~\cite{Alloul20142250}. Subsequently the 
$p p \rightarrow a$ cross section and its decay rates to various channels
were cross checked using the tool \texttt{MadGraph5$\_$aMC@NLO}~\cite{Alwall:2014hca}.

The diphoton rate is very sensitive $k r_c$, precisely due to its exponential dependence 
on the latter. So are the masses of the graviton and gauge boson KK excited states. 
Different values for the parameter $\frac{k}{M_{Pl}}$,
all less than unity, have been chosen while plotting, so that the bulk curvature is 
less than the Planck scale\cite{Davoudiasl:1999jd}. Without this constraint, the classical solution of 5-dimensional Einstein's equation cannot be trusted.

With $k = 0.7 M_{Pl}$ and requiring the first gauge boson KK excitation to be heavier than 3.4 TeV~\cite{ATLAS-CONF-2016-045} ~\footnote{The limit from the non-observation of dilepton peaks, as obtained in \cite{ATLAS-CONF-2016-045}, depends on the
decay width of the heavier vector boson. The limit is as strong as 4.05 TeV for the spin-1 particle 
having SM-like couplings to to fermions, while it could be 3 TeV or lesser with narrower widths. Keeping in mind the fact that a first excited spin-1 KK state has weaker coupling than in the SM, we have taken the limit, somewhat conservatively, as 3.4 TeV. Nonetheless, this leads to a stronger upper limit on $k r_c$, than what can be imposed from KK graviton searches, given the lower bound on the mass of the first graviton KK excited stands at 2.68 TeV~\cite{PhysRevD.90.052005}.}
leads to $k r_c \leq 11.72$. For a given $k r_c$, this upper bound gets tighter upon using a smaller value for $k$.

\begin{figure}[!htbp]
\begin{center}
\includegraphics[scale=0.43]{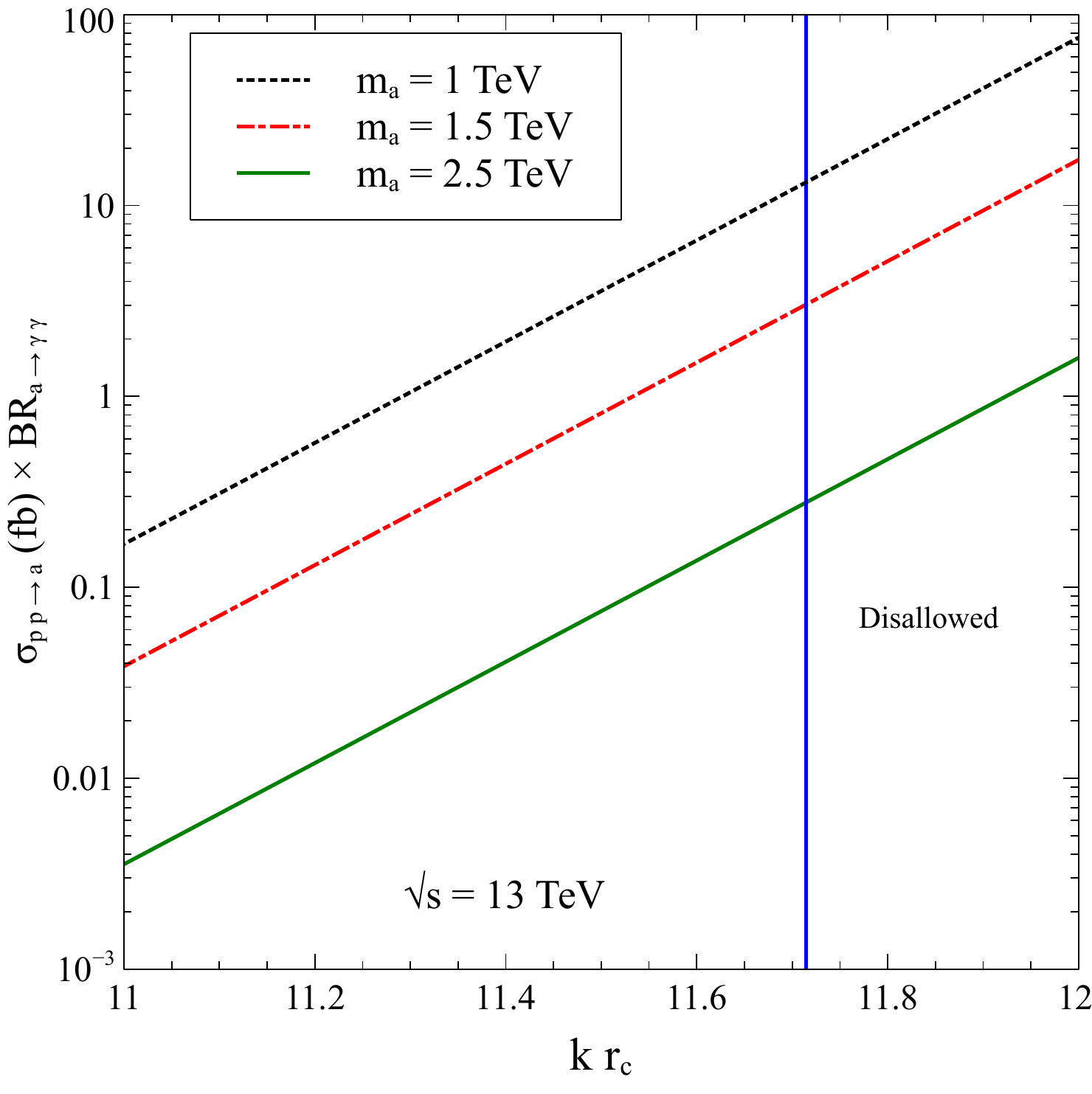}\\
\includegraphics[scale=0.43]{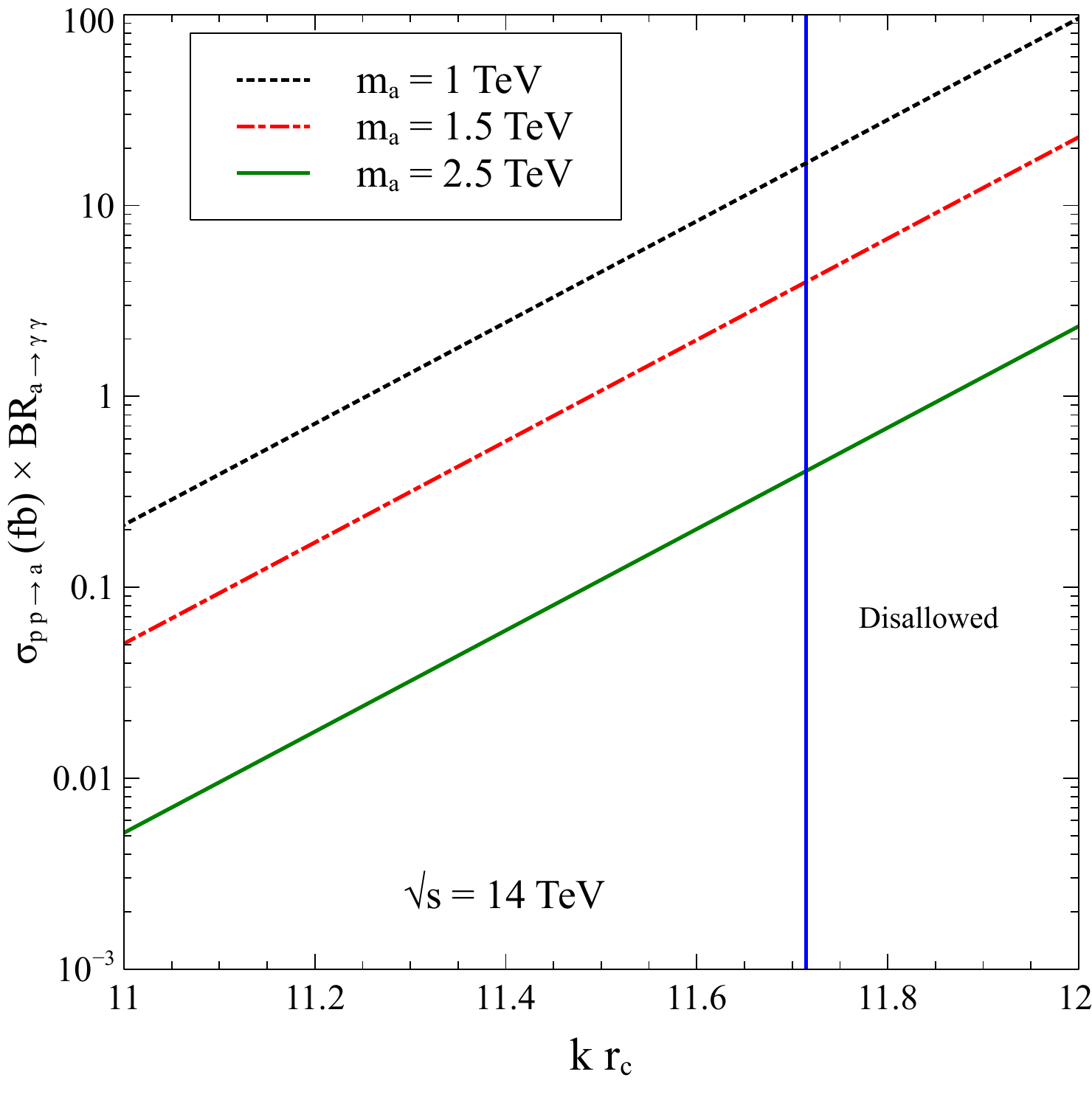}
\caption{Diphoton cross sections at LHC-13 and LHC-14
 for $m_a$ = 1, 1.5 and 2.5 TeV. The region on the right of the
vertical line corresponds to a KK gauge boson below the existing bound.}
\label{f:krc_other}
\end{center}
\end{figure}

The initial results of the  13 TeV collisions have practically
ruled out a diphoton resonance of mass less than 750 GeV that has 'reasonable'
interaction strength with SM particles. For the spin-zero axion considered here,
an exception may occur only if the warp factor $k r_c$ is way below what is 
required for addressing the hierarchy between the Planck and electroweak scales. 
However, higher masses are still within reach. We display the diphoton rates
for axion masses around 1 TeV, 1.5 TeV and 2.5 TeV in Fig.~\ref{f:krc_other}. 
As expected,
the rate goes down as the axion gets heavier. It is seen that a 1 TeV axion 
can have a production cross section of $\simeq$ 5 fb at $\sqrt{s}$ = 13 TeV,
for $k r_c$ = 11.6. This goes up to $\simeq$ 10 fb for $k r_c$ = 11.7. 
Dynamically enhancing
the rate by increasing $k r_c$ to still higher values will invariably come into conflict
with the requirement of heavy KK states. Hence,
to get an appreciable significance, one must wait till the LHC-13 gathers 
more data. The sensitivity however is marginally better for $\sqrt{s} = 14$ TeV. For example,
a 1 TeV axion can yield a $\simeq$ 18 fb cross section for the diphotons in this case.
In principle, the other decay modes (such as digluons) can also
be observed at the LHC. However, the observability of these are perhaps
more challenging than that for the diphotons, since rates of ZZ - peaks undergo branching
fraction suppression, while dijet peaks from gluon pairs are swamped by the background.

\begin{figure}[!htbp]
\begin{center}
\includegraphics[scale=0.43]{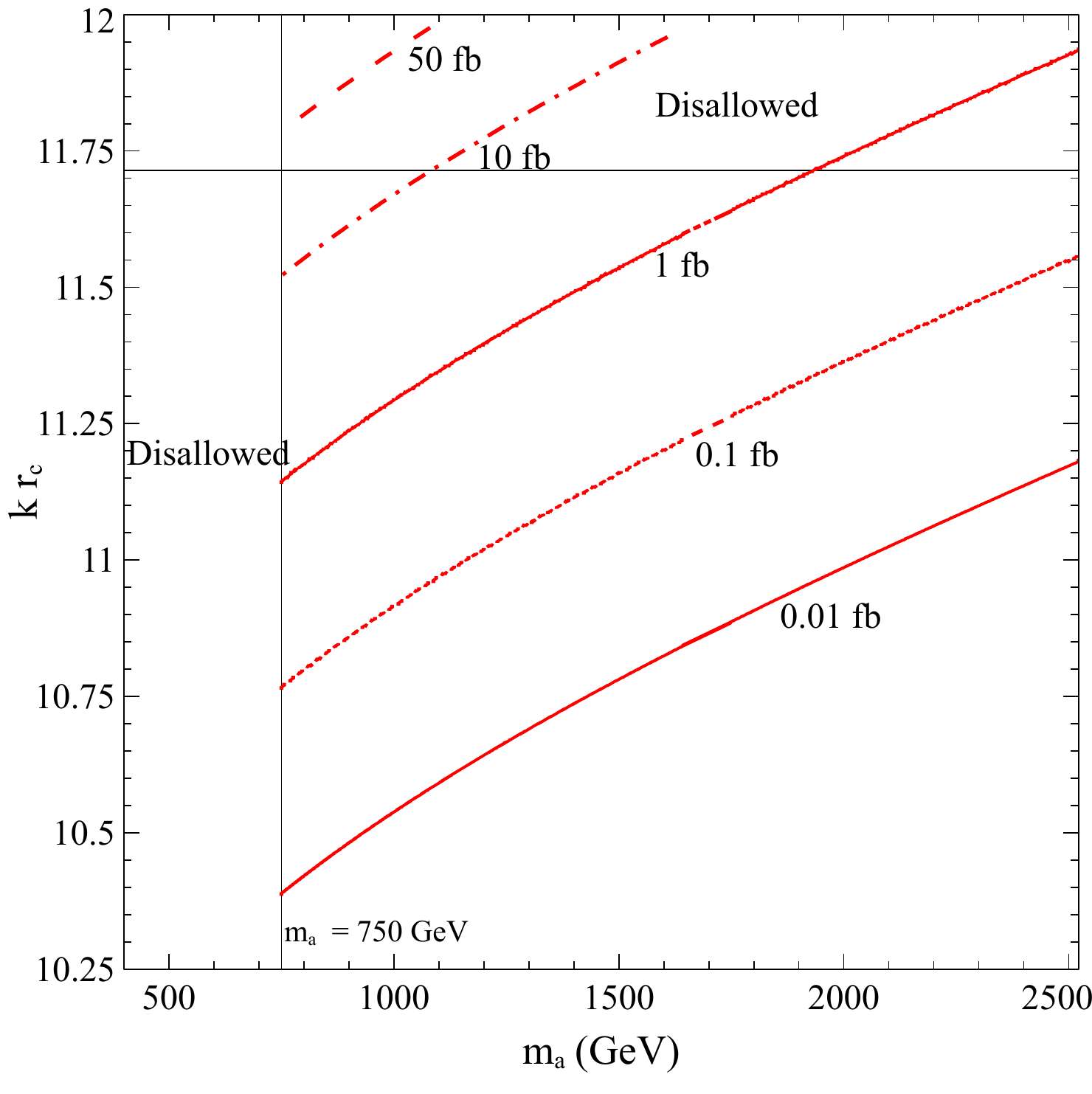}
\caption{Contours of constant cross section in the $m_a$ vs $k r_c$ plane. The region to the
 left of the vertical line is disallowed from the non-observation of a diphoton resonance below
 750 GeV. Similarly, the region above the horizontal line is disallowed from KK gauge boson searches.}
\label{sensitivity}
\end{center}
\end{figure} 

Fig.~\ref{sensitivity} quantitatively depicts the reach of the 13 TeV LHC in detecting
a diphoton resonance that has its origin in the framework under study. 
While a 1 fb diphoton cross section can be predicted even for a 1.9 TeV axion,
a more sizeable rate of 10 fb demands a much lighter axion ($\leq$ 1 TeV). The collider must
gather appreciable luminosity to discern a feeble resonant diphoton rate ($\textless$ 1fb, say)   
from the background. If, upon accumulation of the requisite luminosity, one notices such 'clean' diphoton peaks, the next step would be to see if the $WW$ and $ZZ$ peaks with correlated strengths are also noticeable. In case they are, one has to look further for the presence or absence of corresponding peaks with fermions. If such peaks are absent, then one will be directed to spinless particles which have unsuppressed couplings with gauge boson pairs but no interactions with fermions.
One possible interpretation of such coincidence of observed phenomena may be CS dynamics embedded in a warped geometry.  The exact estimate of the LHC reach for higher masses will require a careful
analysis of cuts and their efficiencies corresponding to the high diphoton invariant mass.

\section{Summary}\label{summary}
In conclusion, a bulk KR field in a 5-dimensional bulk with RS warped geometry
can be connected with an axion in 4-dimensions, which, with a non-perturbatively
acquired mass, can lead to a bump in the diphoton spectrum. 
The most notable feature of this framework is that the production as well as
decay of the axion is triggered by 5-dimensional Chern-Simons terms which are
not {\it ad hoc} introduction but necessitated by the cancellation of gauge anomalies.
Even if one makes the simplifying assumption of universal CS couplings for all gauge bosons,
the diphoton rate for a 1 TeV axion can attain sizeable values at the 13 TeV LHC. 
However, a more realistic estimate is only expected to emerge after detector simulation is
carried out.
We remind that the novelty
of the suggested scenario lies in the fact that the very same the warp factor, which is responsible for the reported diphoton rate, can bridge the hierarchy between the Planck and electroweak 
scales. The visibility of a higher diphoton peak can in principle 
improve in the upcoming 14 TeV runs. 
In all, the final word on a CP-odd scalar around the TeV scale, 
interacting with the gauge bosons through CS terms, will certainly emerge from 
accumulating more experimental data. Nonetheless, we find the above correlation 
rather thought-provoking. 

\section*{Acknowledgement} This work of NC and BM
was partially supported by funding available from the Department of
Atomic Energy, Government of India for the Regional Centre for
Accelerator-based Particle Physics [RECAPP], Harish-Chandra Research Institute.
SSG acknowledges the hospitality of RECAPP while this work was in progress.
We also thank Shankha Banerjee, Subhadeep Mondal and Ashoke Sen 
for helpful discussions.

\bibliography{ref.bib}
\end{document}